# Analysis of Y00 Protocol under Quantum Generalization of a Fast Correlation Attack: Toward Information-Theoretic Security

**Takehisa Iwakoshi**

Department of Information Engineering, Mie University, 1577 Kurimamachiya-cho, Tsu, Mie 514-8507, Japan

Corresponding author: Takehisa Iwakoshi (iwakoshi@cs.info.mie-u.ac.jp).

**ABSTRACT** In our previous work, it was demonstrated that the attacker could not pin-down the correct keys to start the Y00 protocol with a probability of one under the assistance of unlimitedly long known-plaintext attacks and optimal quantum measurements on the attacker's quantum memory. However, there were several assumptions that the Y00 system utilized linear-feedback shift registers as pseudo-random-number generators, and a fast correlation attack was disabled by irregular mapping on the Y00 systems. This study generalizes such an attack to remove the assumptions of the previous work. The framework of the security analyses of this study reiterates two well-known results from the past: (1) Y00 systems would be cryptanalyzed when the system is not designed well; (2) the system is possibly information-theoretically secure when the system is designed well, although the attacker's confidence in the correct key increases over time but the success probability of key recovery does not reach unity in finite time; (3) the breach time of the shared keys is increasingly threatened with time. Hence, a key-refreshment procedure for the Y00 protocol is provided herein. Such security analyses are important not only in key refreshment but also in initial key agreement situations.

**INDEX TERMS** Information-Theoretic Security, Optical network, Quantum cryptography, Quantum detection theory, Secure communications

## 1. INTRODUCTION

Since the first concept of quantum key distribution (QKD) was invented [1], [2], whether information-theoretically secure (ITS) communication is realizable using the laws of quantum physics is a topic that has garnered immense attention.

Around the year 2000, the Y00 protocol (its original name was αη) was proposed by Yuen [3]–[6] for compatibility to existing high-speed and long-distance optical communication infrastructure [3]–[15].

However, the Y00 protocol had been believed to be non-ITS since the fast correlation attack (FCA) on the Y00 protocol was found [16], [17], even after "irregular mapping" was equipped as a countermeasure to the FCA [17], [18]. Hence, the Y00 protocol is believed to be computational secure, while QKDs are said to be ITS.

In our previous work [19], it was shown that the attacker "Eve" could not guess the correct secret keys shared by legitimate users "Alice" and "Bob" with a probability of one even under an unlimitedly long known-plaintext attack (KPA) with the assistance of quantum memory to utilize the quantum and classical multiple-hypotheses testing theory [20]–[22]. The key aspect of the unlimitedly long KPA is to simplify the security analysis of the Y00 protocol because the signals cyclically appear when the effect of the plaintext is subtracted. However, our previous work still assumed that the Y00 system was designed as is; therefore, no security guarantee existed as to how ITS would be realizable against unknown computational attacks.

The purpose of this study is to show that well-designed Y00 systems are immune to the quantum generalization of FCA with the assistance of the unlimitedly long KPA and quantum memories that Eve possesses without any computational assumptions. The analyses in this study demonstrate two results: the main claim of the FCA was recovered against a certain class of Y00 systems not well-designed, while the others would be ITS under the unlimitedly long KPA with the generalized framework of FCA. The framework of the quantum generalization of FCA corresponds to "collective attacks" or "coherent attacks" in the context of QKDs [23], while the existing security analyses of the Y00 protocol were "individual attacks" in the context of QKDs [7]–[18].

The security analyses in this study provide clear security parameters: the security breach time of the Y00 systems, and the minimum error pattern probability that determines the breach time.

This paper is structured as follows. Section II will describe the differences between the conventional stream ciphers and the Y00 protocol in terms of probabilities and information theory to show how the Y00 would be ITS along with the principle of the Y00 protocol. Section III describes the quantum detection theory that Eve performs on her quantum memory storing wire-tapped quantum states. Section IV describes the known concepts of FCA and how it will be generalized in this study. The section also describes the conditions for designing non-ITS Y00 systems and ITS Y00 systems, reiterating known results [16]–[19] in terms of the security breach time. Even if the Y00 system is implemented to be ITS, Eve's success probability increases in the end; hence, Section V describes how the Y00 system securely exchanges fresh keys. Section VI describes the remaining problem, while Section VII states the conclusions.

## 2. BRIEF DESCRIPTION OF PRINCIPLES OF Y00 PROTOCOL

This section describes the differences between the conventional stream ciphers and quantum-noise-randomized stream ciphers, such as the Y00 protocol.

### *A. BRIEF DESCRIPTION OF CONVENTIONAL MATHEMATICAL STREAM CIPHERS*

Let Set($\boldsymbol{V}$) denote the set of possible variables $\boldsymbol{V}$, let |Set($\boldsymbol{V}$)| denote the number of elements in Set($\boldsymbol{V}$). Conventional stream ciphers expand an initial short key $\boldsymbol{k} \in$ Set($\boldsymbol{K}$) into a longer keystream $\boldsymbol{s} \in$ Set($\boldsymbol{S}$) by a pseudo-random-number generator (PRNG). If the KPA is longer than the period of $\boldsymbol{s}$, it completely reveals $\boldsymbol{s}$. Alice sends her message $\boldsymbol{x} \in$ Set($\boldsymbol{X}$) encoded into her ciphertext $\boldsymbol{c} \in$ Set($\boldsymbol{C}$) by

$$\boldsymbol{C} := \boldsymbol{X} + \boldsymbol{S} \bmod 2\,. \tag{1}$$

Then, Eve can recover $\boldsymbol{k}$, irrespective of the complexity of the key expansion algorithm because the key expansion is deterministic, and Eve knows the PRNG, according to Shannon's maxim. In terms of conditional probabilities,

$$\Pr\left(\boldsymbol{S} \mid \boldsymbol{C}, \boldsymbol{X}\right) = \Pr\left(\boldsymbol{K} \mid \boldsymbol{C}, \boldsymbol{X}\right) = 1\,. \tag{2}$$

In terms of Shannon entropy,

$$H\left(\boldsymbol{S} \mid \boldsymbol{C}, \boldsymbol{X}\right) = H\left(\boldsymbol{K} \mid \boldsymbol{C}, \boldsymbol{X}\right) = 0\,. \tag{3}$$

### *B. PRINCIPLE OF THE Y00 PROTOCOL*

To start the Y00 protocol, Alice and Bob must share secret keys, $\boldsymbol{k}$ and $\Delta\boldsymbol{k}$. Then, they expand ($\boldsymbol{k}$, $\Delta\boldsymbol{k}$) $\in$ Set($\boldsymbol{K}$, $\Delta\boldsymbol{K}$) into key streams ($\boldsymbol{s}$, $\Delta\boldsymbol{x}$) $\in$ Set($\boldsymbol{S}$, $\Delta\boldsymbol{X}$) using the common PRNGs equipped in the transmitter and receiver. Subsequently, $\boldsymbol{s}$ is chopped to every $\log_2 M$ bit to form an $M$-ary string $\boldsymbol{s}(t)$ at time slot $t$. A message bit $x(t)$ is encoded into a coherent state $|\alpha[m(t)]\rangle$ as follows:

$$m(t) := \mathrm{Map}\left[\boldsymbol{s}(t)\right] + M\left(\mathrm{Map}\left[\boldsymbol{s}(t)\right] + x(t) + \Delta x(t) \bmod 2\right). \tag{4}$$

Map[$\boldsymbol{s}(t)$] is a projection from $\boldsymbol{s}(t)$ to Map[$\boldsymbol{s}(t)$] $\in \{0, 1, 2, 3, \ldots, M-1\}$. For the detailed characteristics and concrete Map[·], the references [11]–[13] are helpful to understand. Therefore, $x(t) \in \{0, 1\}$ corresponds to a set of quantum states $\{|\alpha[m(t)]\rangle, |\alpha[m(t)+M]\rangle\}$ when Map[$\boldsymbol{s}(t)$] + $\Delta x(t)$ is an even number; otherwise, $\{|\alpha[m(t)+M]\rangle, |\alpha[m(t)]\rangle\}$. In contrast, Bob's receiver sets an optimal threshold to discriminate the set of quantum states based on the shared ($\boldsymbol{s}$, $\Delta\boldsymbol{x}$). Therefore, he decodes $x(t)$ because he knows the value of Map[$\boldsymbol{s}(t)$] + $\Delta x(t)$. Meanwhile, Eve must discriminate the $2M$-ary signals hidden under the overlapping quantum and classical noise because she does not know whether Map[$\boldsymbol{s}(t)$] + $\Delta x(t)$ is even or odd and also does not know $x(t)$.

When Eve can launch KPA longer than $T_{\mathrm{LCM}}$, which is the least common multiple (LCM) of PRNGs' periods in a Y00 system, she will launch an optimal measurement to guess the most probable shared keys. Accordingly, in such a case, the quantum detection theory for multiple-hypothesis testing is required to evaluate the security of the Y00 protocol. Eve obtains the coherent states separated from a beam-splitter as $\rho_{\mathrm{E}}[m(t)]$ and stores the time sequence in her quantum memory. The quantum sequence $\rho_{\mathrm{E}}(\boldsymbol{s}, \Delta\boldsymbol{x}, \boldsymbol{x})$ with the splitting ratio $\eta$ is denoted as follows:

$$\rho_{\mathrm{E}}(\boldsymbol{s}, \Delta\boldsymbol{x}, \boldsymbol{x}) := |\eta\alpha(\boldsymbol{s}, \Delta\boldsymbol{x}, \boldsymbol{x})\rangle\langle\eta\alpha(\boldsymbol{s}, \Delta\boldsymbol{x}, \boldsymbol{x})| = \bigotimes_{t=1}^{T} |\eta\alpha[m(t)]\rangle\langle\eta\alpha[m(t)]|. \tag{5}$$

Note that a set of $(\boldsymbol{s}, \Delta\boldsymbol{x})$ is generated from $(\boldsymbol{k}, \boldsymbol{\Delta k})$. Therefore, Eve needs to solve a $|\mathrm{Set}(\boldsymbol{K},\boldsymbol{\Delta K})|$-ary discrimination problem based on the unlimitedly long KPA like in the case of the conventional stream cipher, although the number of possible signal sequences is $(2M)^T$, which is considerably larger than $|\mathrm{Set}(\boldsymbol{K}, \boldsymbol{\Delta K})|$.

### *C. SECURITY FRAMEWORK OF THE Y00 PROTOCOL*

Shannon proved the necessary condition of perfect secrecy in his Theorem 6 [24] as

$$\Pr(\boldsymbol{X} \mid \boldsymbol{C}) = \Pr(\boldsymbol{C}). \tag{6}$$

After Shannon's perfect secrecy, Wyner showed that almost perfect secrecy could be maintained if the channel to Eve is noisy enough [25]. Such a degradation for Eve is realized only on the physical layer [26]. However, the property of noise on the wire-tap channel is unknown in general situations, especially in the case that Eve has no restrictions on her performance except the laws of physics.

In the case of a Y00 system, the property of noise depends on the implementation of the system. Eve cannot avoid quantum noise in eavesdropping ideal Y00 systems because of overlapping quantum noise caused by the Born rule, as described in Section IV. Consider the following simplified situation. Alice and Bob communicate by a stream cipher

$$\boldsymbol{C} := \boldsymbol{X} + \boldsymbol{S} \bmod 2. \tag{7}$$

In contrast, Eve receives a ciphertext $\boldsymbol{C}_{\mathrm{E}}$ with an error pattern $\boldsymbol{E}$ caused by the noise as

$$\boldsymbol{C}_{\mathrm{E}} := \boldsymbol{X} + \boldsymbol{S} + \boldsymbol{E} \bmod 2. \tag{8}$$

From (8), Eve would be able to recover $\boldsymbol{S}$ if she had known $\boldsymbol{E}$ and $\boldsymbol{X}$ and had observed $\boldsymbol{C}_{\mathrm{E}}$. Hence,

$$\Pr(\boldsymbol{S} \mid \boldsymbol{E}, \boldsymbol{C}_{\mathrm{E}}, \boldsymbol{X}) = 1. \tag{9}$$

However, because Eve never knows $\boldsymbol{E}$,

$$H(\boldsymbol{S} \mid \boldsymbol{C}_{\mathrm{E}}, \boldsymbol{X}) \geq H(\boldsymbol{S}, \boldsymbol{C}_{\mathrm{E}}, \boldsymbol{X}) - H(\boldsymbol{C}_{\mathrm{E}}, \boldsymbol{X}, \boldsymbol{E}) = 0. \tag{10}$$

The equality in (10) holds only when $\boldsymbol{E}$ is a deterministic function of $\boldsymbol{C}_{\mathrm{E}}$ and $\boldsymbol{X}$, which never happens until (8) is satisfied except when Eve can estimate $\boldsymbol{S}$ by algebraic attacks such as FCA, corresponding to the security analysis by equations (6)–(8) in a previous study [6]. The derivation of (10) as well as the reason why the FCA succeeded are given in Appendix A.

Hence, the ideal Y00 protocol never allows Eve to obtain $\boldsymbol{S}$ as well as $\boldsymbol{K}$ deterministically, which is significantly different from conventional stream ciphers in Section II.A. Therefore, (10) suggests that an ideal Y00 system is ITS.

Accurately, von Neumann entropy is more suitable than Shannon entropy because Eve is supposed to store wire-tapped quantum states in her quantum memory. However, Eve must measure her memory to obtain the most likely results. Therefore, Shannon entropy is sufficient because the measured results are classical. The above point will be discussed in Section IV.E.

### *D. OTHER CLASSES OF ATTACKS*

The readers may wonder why this study treats only KPA while there are several classes of attacks as follows.

1. Ciphertext only attacks (COA); The attacker utilizes the only ciphertext to obtain the plaintext or the key.
2. Known plaintext attacks (KPA); The attacker knows the plaintext then tries to find the encryption key.
3. Chosen plaintext attacks (CPA); The attacker can access the encryption system to obtain the pair of a known plaintext and the corresponding ciphertext.
4. Chosen ciphertext attacks (CCA); The attacker can access the decryption system, then injects ciphertext to obtain the corresponding plaintext.

In any classes except COA, the attacker can obtain the pair of the plaintext and the corresponding ciphertext to perform key-recovery attacks in the Y00 systems. Therefore, there is no significant difference between unlimitedly long KPA in this study and other cryptologic attack classes.

## 3. BRIEF DESCRIPTION OF QUANTUM DETECTION THEORY

This section describes how Eve utilizes her quantum memory and performs the optimal measurement. The description shows that Eve's success probability in obtaining the correct keys never reaches unity.

### *A. BRIEF DESCRIPTION OF THE QUANTUM DETECTION THEORY*

From this section onward, $(\boldsymbol{s}, \Delta\boldsymbol{x}) \in \mathrm{Set}(\boldsymbol{S}, \Delta\boldsymbol{X})$ is abbreviated as $\boldsymbol{r} \in \mathrm{Set}(\boldsymbol{R})$ for simplicity. In the quantum detection theory, $W(\boldsymbol{r},\boldsymbol{x})$ is a Hermitian risk operator, and $\{M_{\mathrm{E}}(\boldsymbol{r}\,|\,\boldsymbol{x})\}$ is a set of Eve's optimal measurement operators to minimize her average error rate conditioned on the known $\boldsymbol{x}$ [20], [21]. The necessary-and-sufficient conditions of Eve's optimum $\{M_{\mathrm{E}}(\boldsymbol{r}\,|\,\boldsymbol{x})\}$ in (11) are described by (12)–(17).

$$M_{\mathrm{E}}(\boldsymbol{r}\,|\,\boldsymbol{x}) := \left|(\boldsymbol{r}\,|\,\boldsymbol{x})\right\rangle\left\langle(\boldsymbol{r}\,|\,\boldsymbol{x})\right|. \tag{11}$$

$$\sum\nolimits_{\boldsymbol{r}\in\mathrm{Set}(\boldsymbol{R})} M_{\mathrm{E}}\left(\boldsymbol{r}\,|\,\boldsymbol{x}\right) = I. \tag{12}$$

$$W\left(\boldsymbol{r},\boldsymbol{x}\right) := -\sum\nolimits_{\boldsymbol{r}'\in\mathrm{Set}(\boldsymbol{R})} \delta_{\boldsymbol{r},\boldsymbol{r}'}\,\Pr(\boldsymbol{r}')\rho_{\mathrm{E}}(\boldsymbol{r}',\boldsymbol{x}) = -\Pr(\boldsymbol{r})\left|\eta\alpha(\boldsymbol{r},\boldsymbol{x})\right\rangle\left\langle\eta\alpha(\boldsymbol{r},\boldsymbol{x})\right|. \tag{13}$$

$$\Gamma(\boldsymbol{x}) := \sum\nolimits_{\boldsymbol{r}\in\mathrm{Set}(\boldsymbol{R})} M_{\mathrm{E}}\left(\boldsymbol{r}\,|\,\boldsymbol{x}\right)W\left(\boldsymbol{r},\boldsymbol{x}\right) = \sum\nolimits_{\boldsymbol{r}\in\mathrm{Set}(\boldsymbol{R})} W\left(\boldsymbol{r},\boldsymbol{x}\right)M_{\mathrm{E}}\left(\boldsymbol{r}\,|\,\boldsymbol{x}\right). \tag{14}$$

$$\left[W\left(\boldsymbol{r},\boldsymbol{x}\right) - \Gamma(\boldsymbol{x})\right]M_{\mathrm{E}}\left(\boldsymbol{r}\,|\,\boldsymbol{x}\right) = M_{\mathrm{E}}\left(\boldsymbol{r}\,|\,\boldsymbol{x}\right)\left[W\left(\boldsymbol{r},\boldsymbol{x}\right) - \Gamma(\boldsymbol{x})\right] = \mathbf{0}. \tag{15}$$

$$M_{\mathrm{E}}\left(\boldsymbol{r}\,|\,\boldsymbol{x}\right)\left[W\left(\boldsymbol{r}',\boldsymbol{x}\right) - W\left(\boldsymbol{r},\boldsymbol{x}\right)\right]M_{\mathrm{E}}\left(\boldsymbol{r}'\,|\,\boldsymbol{x}\right) = \mathbf{0}. \tag{16}$$

$$W\left(\boldsymbol{r},\boldsymbol{x}\right) - \Gamma(\boldsymbol{x}) \geq 0. \tag{17}$$

Once $\{M_{\mathrm{E}}(\boldsymbol{r}\,|\,\boldsymbol{x})\}$ is determined, from the Cauchy–Schwarz inequality, Eve's average success probability with known $\boldsymbol{x}$ denoted by $-\mathrm{tr}\,\Gamma(\boldsymbol{x})$ is maximized as follows [19].

$$\left[-\mathrm{tr}\,\Gamma(\boldsymbol{x})\right]^2 = \left[\sum\nolimits_{\boldsymbol{r}\in\mathrm{Set}(\boldsymbol{R})} \Pr(\boldsymbol{r})\left|\left\langle\eta\alpha(\boldsymbol{r},\boldsymbol{x})\middle|(\boldsymbol{r}\,|\,\boldsymbol{x})\right\rangle\right|^2\right]^2 \leq \left[\sum\nolimits_{\boldsymbol{r}\in\mathrm{Set}(\boldsymbol{R})} \Pr(\boldsymbol{r})^2\right]\left[\sum\nolimits_{\boldsymbol{r}\in\mathrm{Set}(\boldsymbol{R})} \left|\left\langle\eta\alpha(\boldsymbol{r},\boldsymbol{x})\middle|(\boldsymbol{r}\,|\,\boldsymbol{x})\right\rangle\right|^4\right]. \tag{18}$$

The equality is satisfied when Eve can choose $\{M_{\mathrm{E}}(\boldsymbol{r}\,|\,\boldsymbol{x})\}$ so that it satisfies (19).

$$\Pr(\boldsymbol{r}) = \frac{\left|\left\langle\eta\alpha(\boldsymbol{r},\boldsymbol{x})\middle|(\boldsymbol{r}\,|\,\boldsymbol{x})\right\rangle\right|^2}{\sum\nolimits_{\boldsymbol{r}\in\mathrm{Set}(\boldsymbol{R})} \left|\left\langle\eta\alpha(\boldsymbol{r},\boldsymbol{x})\middle|(\boldsymbol{r}\,|\,\boldsymbol{x})\right\rangle\right|^2}. \tag{19}$$

Hence, Eve's average success probability (20) is satisfied if and only if (21) is satisfied.

$$\max\left[-\mathrm{tr}\,\Gamma(\boldsymbol{x})\right] = \frac{\sum\nolimits_{\boldsymbol{r}\in\mathrm{Set}(\boldsymbol{R})} \left|\left\langle\eta\alpha(\boldsymbol{r},\boldsymbol{x})\middle|(\boldsymbol{r}\,|\,\boldsymbol{x})\right\rangle\right|^4}{\sum\nolimits_{\boldsymbol{r}\in\mathrm{Set}(\boldsymbol{R})} \left|\left\langle\eta\alpha(\boldsymbol{r},\boldsymbol{x})\middle|(\boldsymbol{r}\,|\,\boldsymbol{x})\right\rangle\right|^2} < 1. \tag{20}$$

$$\left|\left\langle\eta\alpha(\boldsymbol{r},\boldsymbol{x})\middle|(\boldsymbol{r}\,|\,\boldsymbol{x})\right\rangle\right|^2 < 1. \tag{21}$$

### *B. QUANTUM DETECTION FOR SEQUENTIAL COHERENT SIGNALS*

To provide detailed analyses for the sequential Y00 signals, the over-completeness property of coherent states in (22) is required with $D(\mathrm{All})$ covering the entire complex plains. In (23), $D(\boldsymbol{r} \mid \boldsymbol{x})$ is an integration domain for signals originated from $(\boldsymbol{r}, \boldsymbol{x})$, satisfying (24)–(26).

$$\bigotimes_{t=1}^{T} \int_{\alpha(t) \in D(\mathrm{All})} \pi^{-1} |\alpha(t)\rangle\langle\alpha(t)| d\alpha(t) = I . \tag{22}$$

$$M_{\mathrm{E}}(\boldsymbol{r}+\boldsymbol{e} \mid \boldsymbol{x}) := \bigotimes_{t=1}^{T} \int_{\alpha(t) \in D(\boldsymbol{r}+\boldsymbol{e} \mid \boldsymbol{x})} \pi^{-1} |\alpha(t)\rangle\langle\alpha(t)| d\alpha(t) . \tag{23}$$

$$\bigcup_{\boldsymbol{e} \in \mathrm{Set}[\boldsymbol{E}(\boldsymbol{r} \mid \boldsymbol{x})]} D(\boldsymbol{r}+\boldsymbol{e} \mid \boldsymbol{x}) = D(\boldsymbol{r} \mid \boldsymbol{x}) . \tag{24}$$

$$\bigcup_{\boldsymbol{r} \in \mathrm{Set}(\boldsymbol{R})} D(\boldsymbol{r} \mid \boldsymbol{x}) = D(\mathrm{All}) . \tag{25}$$

$$D(\boldsymbol{r}+\boldsymbol{e} \mid \boldsymbol{x}) \cap D(\boldsymbol{r}'+\boldsymbol{e}' \neq \boldsymbol{r}+\boldsymbol{e} \mid \boldsymbol{x}) = \varnothing . \tag{26}$$

$$M_{\mathrm{E}}(\boldsymbol{r} \mid \boldsymbol{x}) := \sum_{\boldsymbol{e} \in \mathrm{Set}[\boldsymbol{E}(\boldsymbol{r} \mid \boldsymbol{x})]} M_{\mathrm{E}}(\boldsymbol{r}+\boldsymbol{e} \mid \boldsymbol{x}) . \tag{27}$$

Here, $\boldsymbol{e}$ is an error pattern from the correct $\boldsymbol{r}$ as a result of Eve's measurement operator $M_{\mathrm{E}}(\boldsymbol{r}+\boldsymbol{e} \mid \boldsymbol{x})$ for the convenience in security analyses discussed in Section IV. $\mathrm{Set}[\boldsymbol{E}(\boldsymbol{r} \mid \boldsymbol{x})]$ is a set of error patterns $\boldsymbol{e}$ originated from $\boldsymbol{r}$ conditioned on $\boldsymbol{x}$.

## 4. QUANTUM GENERALIZATION OF FCA AND SECURITY REQUIREMENTS

Our previous study [19] showed that the Y00 protocol would be secure against an unlimitedly long KPA under several assumptions. This section removes the assumptions to generalize FCA to evaluate the security of the Y00 protocol. Thus, the disadvantage of the Y00 protocol compared with that of the QKD + One-Time Pad (OTP) described in Section IV.A. of [27] would be removed.

### *A. BASIC CONCEPT OF FCA AND GENERALIZATION*

The fundamental concept of the original FCA is as follows [16], [17].

1. Unless Map[·] is designed well, some bits in $\boldsymbol{r}(t)$ are not sufficiently hidden in quantum noise.
2. Hence, it reveals some bits in the keystream from the linear feedback shift register (LFSR) to Eve.
3. Eve calculates the most likely seed key from the revealed bits in the keystream; some erroneous bits are even corrected by applying error-correction code.

Fig. 3 of the literature [16] showed the above situation. Then, the literature [17] formulated the attack scheme. However, if Map[·] is well-designed, quantum noise hides all bits in $\boldsymbol{r}(t)$ almost equally, as proposed by the literature [17] and [18]. A numerical simulation is performed in the literature [18].

However, note that the above countermeasure is for a specific attack against a certain Y00 implementation. There should be more general attacks.

To construct a more general attack, the above assumptions listed in the literature [19] must be removed as follows.

1. A Y00 system employs arbitrary PRNGs to expand the shared secret keys into $\boldsymbol{r}$.
2. Map[·] is not specified; however, some of the bits in $\boldsymbol{r}(t)$ may not be covered by quantum noise.
3. Eve guesses the most likely $\boldsymbol{r}$ by a collective measurement on her quantum memory, including known plaintext, which is different from an individual measurement on each signal in the original FCA.

All analyses are performed similarly to the discussions in Section II.

### *B. DETAILED DESCRIPTION OF GENERALIZED ATTACK*

Let us denote the duration of KPA as $T = N \cdot T_{\mathrm{LCM}}$ and the number of error patterns $\boldsymbol{e}$ as $n(\boldsymbol{e}|\,\boldsymbol{x}, \boldsymbol{r})$ during $T$. Then,

$$\sum\nolimits_{\boldsymbol{e}\in\{0,1\}^{|\boldsymbol{e}|}} n(\boldsymbol{e}\,|\,\boldsymbol{r},\boldsymbol{x}) = N\,. \tag{28}$$

Here, $|\boldsymbol{e}| = T_{\mathrm{LCM}} \log_2 (2M)$ is the length of the error patterns $\boldsymbol{e}$. Then, Eve's success probability $\Pr(\boldsymbol{r}\,|\,\boldsymbol{r},\boldsymbol{x})$ is given as

$$1 \ge \Pr\left(\boldsymbol{r}\,|\,\boldsymbol{r},\boldsymbol{x}\right) = \sum\nolimits_{\{n(\boldsymbol{e}|\boldsymbol{r},\boldsymbol{x})\}\in\Omega(\boldsymbol{r}|\boldsymbol{x})} \left[N!\right] \prod\nolimits_{\boldsymbol{e}\in\{0,1\}^{|\boldsymbol{e}|}} \left[n(\boldsymbol{e}\,|\,\boldsymbol{r},\boldsymbol{x})!\right]^{-1} \Pr(\boldsymbol{e}\,|\,\boldsymbol{r},\boldsymbol{x})^{n(\boldsymbol{e}|\boldsymbol{r},\boldsymbol{x})}\,. \tag{29}$$

Here, $\Omega(\boldsymbol{r}\,|\,\boldsymbol{x})$ is a set of $\{n(\boldsymbol{e}\,|\,\boldsymbol{r},\boldsymbol{x})\}$ whereby the detected state originates from $\boldsymbol{r}$ under known $\boldsymbol{x}$, and its complemental set is $\Omega(\boldsymbol{r}\,|\,\boldsymbol{x})^{\mathrm{C}} := \Omega(\mathrm{All}) - \Omega(\boldsymbol{r}\,|\,\boldsymbol{x})$. The upper bound of $\Pr(\boldsymbol{r}\,|\,\boldsymbol{r},\boldsymbol{x})$ is

$$\Pr\left(\boldsymbol{r}\,|\,\boldsymbol{r},\boldsymbol{x}\right) \le 1 - \left[1 - \Pr(\boldsymbol{r})\right] \exp\left[-\left(N/N_{\mathrm{Breach}}\right)\ln 2\right]. \tag{30}$$

Here, $N_{\mathrm{Breach}}$ is defined as follows:

$$1/N_{\mathrm{Breach}} := -\log_2\left[(2M)^{T_{\mathrm{LCM}}} \min\nolimits_{\boldsymbol{e}\in\{0,1\}^{|\boldsymbol{e}|}} \Pr(\boldsymbol{e}\,|\,\boldsymbol{r},\boldsymbol{x})\right]. \tag{31}$$

$$\min\nolimits_{\boldsymbol{e}\in\{0,1\}^{|\boldsymbol{e}|}} \Pr(\boldsymbol{e}\,|\,\boldsymbol{r},\boldsymbol{x}) \le (2M)^{-T_{\mathrm{LCM}}}\,. \tag{32}$$

The derivations for (29)–(31) are described in Appendix B. The relation between $\Pr(\boldsymbol{r}\,|\,\boldsymbol{r},\boldsymbol{x})$ and $-\mathrm{tr}\,\Gamma(\boldsymbol{x})$ is given by

$$-\mathrm{tr}\,\Gamma(\boldsymbol{x}) = \sum\nolimits_{\boldsymbol{r}\in\mathrm{Set}(\boldsymbol{R})} \Pr(\boldsymbol{r}) \Pr\left(\boldsymbol{r}\,|\,\boldsymbol{r},\boldsymbol{x}\right). \tag{33}$$

### *C. NON-ITS Y00 SYSTEMS*

If $1 > N_{\mathrm{Breach}} \to 0$, then the Y00 system reaches $\Pr(\boldsymbol{r} \mid \boldsymbol{r}, \boldsymbol{x}) \to 1$ immediately after the protocol begins. Such a condition is satisfied when

$$1/2 > (2M)^{T_{\mathrm{LCM}}} \min_{\boldsymbol{e} \in \{0,1\}^{|\boldsymbol{e}|}} \Pr(\boldsymbol{e} \mid \boldsymbol{r}, \boldsymbol{x}) \to 0 . \tag{34}$$

Therefore, in the generalized framework of FCA in this study, the known results [16], [17] are obtained under the condition (34). Such Y00 systems cannot be ITS irrespective of the complexity of PRNGs' algorithms, as discussed in Section II.

As a conclusion of this section, the generalization of FCA on Y00 systems is given in terms of the information theory and probabilities under Eve's optimal quantum measurement without any computational assumptions. It was shown that there exist some non-ITS Y00 systems if their implementations are invalid.

### *D. REQUIREMENTS ON ITS Y00 SYSTEMS*

To implement an ITS Y00 system, the requirement is $\infty > N_{\mathrm{Breach}} \geq 1$. Therefore,

$$1 > (2M)^{T_{\mathrm{LCM}}} \min_{\boldsymbol{e} \in \{0,1\}^{|\boldsymbol{e}|}} \Pr(\boldsymbol{e} \mid \boldsymbol{r}, \boldsymbol{x}) \geq 1/2 . \tag{35}$$

If (35) is satisfied, then the right-hand side of (30) never reaches unity while $\infty > N \geq 0$, which is unlike the conventional stream ciphers described in Section II.A, although Eve's success probability asymptotically increases as time $T$ increases. Therefore, the generalization framework of FCA in this study again recovers the known result [19].

Moreover, the following simple case satisfies the condition.

$$\Pr\left(\boldsymbol{r} \mid \boldsymbol{r}, \boldsymbol{x}\right) = \Pr(\boldsymbol{r}) . \tag{36}$$

when

$$(2M)^{T_{\mathrm{LCM}}} \min_{\boldsymbol{e} \in \{0,1\}^{|\boldsymbol{e}|}} \Pr(\boldsymbol{e} \mid \boldsymbol{r}, \boldsymbol{x}) = 1 . \tag{37}$$

The above result means that irrespective of how long Eve launches KPA, her guessing probability remains the same as her pure guessing probability because of the infinitely long $N_{\mathrm{Breach}}$, although the condition may not be satisfied.

The above conclusions are an analogy of OTP with a non-IID key string. If the key string is far from IID, its statistical property may give Eve a hint on the plaintext corresponding to $\boldsymbol{r}$ in this work because of the absence of (6).

### *E. EFFECT OF EVE'S LOCAL OPERATIONS*

This section discusses whether Eve can obtain any advantage by her local quantum operations, including any trace-preserving completely positive (TPCP) maps in her quantum memory. Such quantum operations include classical operations, as well. Hence, her optical amplifications on the stolen signals are theoretically included.

Without her local operations, her optimal measurement is given by (18), which is given here again.

$$-\operatorname{tr}\Gamma(\boldsymbol{x}) = \sum_{\boldsymbol{r}\in\mathrm{Set}(\boldsymbol{R})} \Pr(\boldsymbol{r}) \left|\langle \eta\alpha(\boldsymbol{x},\boldsymbol{r}) | (\boldsymbol{r}\,|\,\boldsymbol{x})\rangle\right|^2 = \sum_{\boldsymbol{r}\in\mathrm{Set}(\boldsymbol{R})} \Pr(\boldsymbol{r}) \operatorname{tr}\left[M_{\mathrm{E}}(\boldsymbol{r}\,|\,\boldsymbol{x})\rho_{\mathrm{E}}(\boldsymbol{r},\boldsymbol{x})\right]. \tag{38}$$

When she performs a TPCP operation $\Lambda_{\mathrm{E}}$ on her system, her optimal measurement operator is denoted as $\{M'_{\mathrm{E}}(\boldsymbol{r}\,|\,\boldsymbol{x})\}$ ; hence, her success probability is $-\operatorname{tr}\Gamma'(\boldsymbol{x})$ .

$$-\operatorname{tr}\Gamma'(\boldsymbol{x}) = \sum_{\boldsymbol{r}\in\mathrm{Set}(\boldsymbol{R})} \Pr(\boldsymbol{r}) \operatorname{tr}\left[M'_{\mathrm{E}}(\boldsymbol{r}\,|\,\boldsymbol{x})\Lambda_{\mathrm{E}}\left(\rho_{\mathrm{E}}(\boldsymbol{r},\boldsymbol{x})\right)\right]. \tag{39}$$

Therefore, the problem is whether $-\operatorname{tr}\Gamma'(\boldsymbol{x}) > -\operatorname{tr}\Gamma(\boldsymbol{x})$ or not.

Such a $\Lambda_{\mathrm{E}}$ is rewritten by a unitary operator $U_{\mathrm{EQ}}$ by adding a virtual ancilla Q to Eve's system.

$$\Lambda_{\mathrm{E}}\left(\rho_{\mathrm{E}}(\boldsymbol{r},\boldsymbol{x})\right) := \operatorname{tr}_{\mathrm{Q}}\left[U_{\mathrm{EQ}}\left(\rho_{\mathrm{E}}(\boldsymbol{r},\boldsymbol{x}) \otimes |0\rangle\langle 0|_{\mathrm{Q}}\right)U_{\mathrm{EQ}}^{\dagger}\right]. \tag{40}$$

Then,

$$-\operatorname{tr}\Gamma'(\boldsymbol{x}) = \sum_{\boldsymbol{r}\in\mathrm{Set}(\boldsymbol{R})} \Pr(\boldsymbol{r}) \operatorname{tr}\left[M''_{\mathrm{EQ}}(\boldsymbol{r}\,|\,\boldsymbol{x})\rho_{\mathrm{EQ}}(\boldsymbol{r},\boldsymbol{x})\right]. \tag{41}$$

$$M''_{\mathrm{EQ}}(\boldsymbol{r}\,|\,\boldsymbol{x}) := U_{\mathrm{EQ}}^{\dagger} M'_{\mathrm{EQ}}(\boldsymbol{r}\,|\,\boldsymbol{x}) U_{\mathrm{EQ}}. \tag{42}$$

$$\rho_{\mathrm{EQ}}(\boldsymbol{r},\boldsymbol{x}) := \rho_{\mathrm{E}}(\boldsymbol{r},\boldsymbol{x}) \otimes |0\rangle\langle 0|_{\mathrm{Q}}. \tag{43}$$

If (44) is satisfied $\{M''_{\mathrm{E}}(\boldsymbol{r}\,|\,\boldsymbol{x})\}$ is an optimal measurement satisfying (12)–(17),

$$\operatorname{tr}_{\mathrm{E}}\left[M''_{\mathrm{EQ}}(\boldsymbol{r}\,|\,\boldsymbol{x})\rho_{\mathrm{EQ}}(\boldsymbol{r},\boldsymbol{x})\right] = M_{\mathrm{E}}(\boldsymbol{r}\,|\,\boldsymbol{x})\rho_{\mathrm{E}}(\boldsymbol{r},\boldsymbol{x}). \tag{44}$$

Otherwise, $-\operatorname{tr}\Gamma'(\boldsymbol{x}) < -\operatorname{tr}\Gamma(\boldsymbol{x})$. Therefore, her local TPCP operations never give her any advantages.

The above conclusion sounds natural since the Holevo quantity bounds the accessible information, while the quantum data processing inequality described by von Neumann entropy tells degradation of the obtainable information by TPCP maps.

### *F. NUMERICAL EXAMPLE*

FIGURE 1(a) shows characteristics of (31) while 1(b) shows examples of (30) with parameters in its caption.

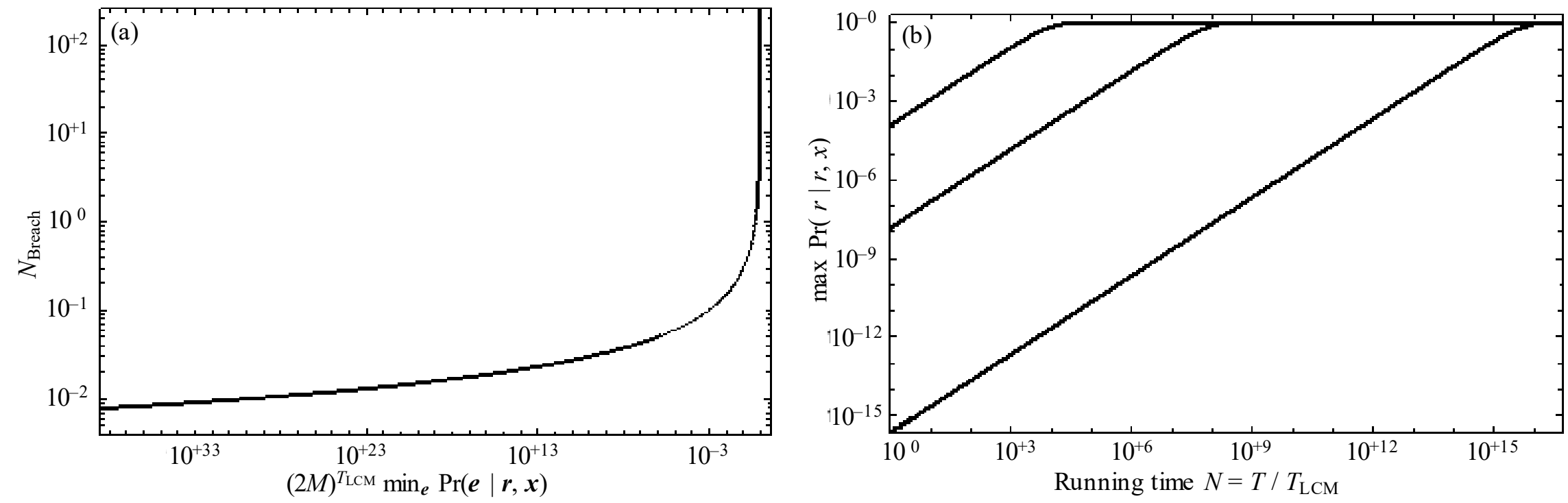

**FIGURE 1. (a) $N_{\text{Breach}}$ vs. $\min_e$ P($e$ | $r$, $x$) in (31), (b) Eve's success probability with parameters in (30) of Pr($r$) = $|\text{Set}(\boldsymbol{R})|^{-1} = (2^{256} - 1)^{-2}$, $1/N_{\text{Breach}} = 1 - 2^{-13}, 1 - 2^{-26}, 1 - 2^{-52}$ from the top curve.**

With regard to the time scale of $N_{\text{Breach}}$, her success probability reaches almost unity. Because her success probability must be sufficiently suppressed, the legitimate users must set a security threshold $P_{\text{Th}}$ to a certain level, and then, they must estimate the actual breach time of the shared key. However, $N_{\text{Breach}}$ is very sensitive to $\min_e \Pr(\boldsymbol{e} \mid \boldsymbol{r}, \boldsymbol{x})$ when the users need sufficiently long $N_{\text{Breach}}$. Hence, in the actual situation, parameters must be carefully estimated from the designs of the corresponding Y00 systems.

### *G. POSSIBLY BETTER IMPLEMENTATIONS*

To realize a condition close to the ideal situation (36), the simple implementation described in Section II.B. may not be sufficient.

A possible solution is to add a classical randomization technique named deliberate signal randomization (DSR). Security enhancement by the technique originated was obtained from a previous study [3]. Then, implementation using an additional PRNG was proposed [8], [11], called the "keyed DSR." The concept of DSR is a modification of (4).

$$m(t) := \text{Map}\left[\boldsymbol{s}(t) + \boldsymbol{d}(t) \bmod 2\right] + M\left(\text{Map}\left[\boldsymbol{s}(t)\right] + x(t) + \Delta x(t) \bmod 2\right). \tag{45}$$

Here, $\boldsymbol{d}(t)$ is a random string of length $|\boldsymbol{d}(t)| = |\boldsymbol{s}(t)|$. If $\boldsymbol{d}(t)$ has a probability distribution of independent and identically distributed (IID), arbitral error patterns $\boldsymbol{e}$ will occur with the same probability. However, the keyed DSR will not satisfy the IID condition because $\boldsymbol{d}(t)$ is deterministic. It would enlarge the key-space of Y00 systems and give a longer $T_{\text{LCM}}$; however, it will never be an essential solution because the output of PRNG is periodic. A true-random DSR is recommended. Another possible solution is to scramble Map[·] by an additional PRNG as Map($t$)[·].

PRNGs in Y00 systems must be chosen carefully. Recall that the Y00 systems described in this study consist of at least two PRNGs; one is for $\boldsymbol{s}$ to select the signal level, and the other is for $\Delta \boldsymbol{x}$ to scramble the plaintext $\boldsymbol{x}$. It is well-known that LFSRs do not even have a statistically good property. A combination of several LFSRs shows correlations between them. At least statistically good PRNGs must be chosen, such as Mersenne Twister [28] or TinyMT [29].

### *H. EFFECT OF TRUE-RANDOM DSR*

By a true-random DSR, pure states being sent from Alice become mixed states for Eve, as follows:

$$\rho_{\mathrm{E}}(\boldsymbol{r},\boldsymbol{x}) := \sum_{\boldsymbol{d}\in\{0,1\}^{|\boldsymbol{d}|}} \Pr(\boldsymbol{d}) \left|\eta\alpha(\boldsymbol{r},\boldsymbol{x},\boldsymbol{d})\right\rangle\left\langle\eta\alpha(\boldsymbol{r},\boldsymbol{x},\boldsymbol{d})\right| . \tag{46}$$

If $\Pr(\boldsymbol{d})$ is uniform, $\Pr(\boldsymbol{d}) = M^{-T}$. Thus, Eve's success probability $-\mathrm{tr}\,\Gamma'(\boldsymbol{x})$ is given by her corresponding optimal measurement $\{M'_{\mathrm{E}}(\boldsymbol{r}\,|\,\boldsymbol{x})\}$ as follows:

$$-\mathrm{tr}\,\Gamma'(\boldsymbol{x}) = \sum_{\boldsymbol{r}\in\mathrm{Set}(\boldsymbol{R})} \Pr(\boldsymbol{r})\,\mathrm{tr}\left[M'_{\mathrm{E}}(\boldsymbol{r}\,|\,\boldsymbol{x})\rho_{\mathrm{E}}(\boldsymbol{r},\boldsymbol{x})\right]. \tag{47}$$

The above situation is similar to the situation in Section IV.E, in which Eve performs her local TPCP operations, resulting in her local quantum system becoming a mixed state. Hence, by a similar procedure,

$$-\mathrm{tr}\,\Gamma'(\boldsymbol{x}) = \sum_{\boldsymbol{r}\in\mathrm{Set}(\boldsymbol{R})} \Pr(\boldsymbol{r})\,\mathrm{tr}\left[M''_{\mathrm{EQ}}(\boldsymbol{r}\,|\,\boldsymbol{x})\rho_{\mathrm{EQ}}(\boldsymbol{r},\boldsymbol{x})\right]. \tag{48}$$

$$M''_{\mathrm{EQ}}(\boldsymbol{r}\,|\,\boldsymbol{x}) := U^{\dagger}_{\mathrm{EQ}} M'_{\mathrm{EQ}}(\boldsymbol{r}\,|\,\boldsymbol{x}) U_{\mathrm{EQ}} . \tag{49}$$

$$\rho_{\mathrm{EQ}}(\boldsymbol{r},\boldsymbol{x}) := \rho_{\mathrm{E}}(\boldsymbol{r},\boldsymbol{x}) \otimes \left|0\right\rangle\left\langle 0\right|_{\mathrm{Q}} . \tag{50}$$

Because Section IV.E concluded that Eve's local TPCP operations never give her any advantages by the same discussion, the conclusion is $-\mathrm{tr}\,\Gamma'(\boldsymbol{x}) \le -\mathrm{tr}\,\Gamma(\boldsymbol{x})$, as well.

## 5. KEY-REFRESHMENT BY LEFTOVER HASHING IN QUANTUM NOISE

Section IV.D showed that Y00 systems would be ITS if their implementations are appropriate. However, Eve becomes confident regarding the correct keys over time. Therefore, this section provides a method to refresh the shared keys between Alice and Bob before the Y00 systems are threatened.

### *A. LEFTOVER HASH LEMMA*

To share a set of fresh keys, Alice or Bob sends a random string $\boldsymbol{x} \in \mathrm{Set}(\boldsymbol{X})$ instead of their messages, where $\boldsymbol{x}$ is an error correction code containing a hash function $\boldsymbol{h} \in \mathrm{Set}(\boldsymbol{H})$, and a seed key $\boldsymbol{k}_{\mathrm{R}} \in \mathrm{Set}(\boldsymbol{K}_{\mathrm{R}})$ to generate ($\boldsymbol{k}_{\mathrm{new}}$, $\Delta\boldsymbol{k}_{\mathrm{new}}$) as follows.

$$(\boldsymbol{k}_{\mathrm{New}}, \Delta\boldsymbol{k}_{\mathrm{New}}) = \boldsymbol{h}(\boldsymbol{k}_{\mathrm{R}}) . \tag{51}$$

Because Eve never knows $\boldsymbol{X}$, her attack is now limited to ciphertext only attacks (COA).

According to the ordinary leftover hash lemma (LHL) [30], [31] there exists a strong ($\tau$, $\kappa$, $\varepsilon$)-randomness extractor to obtain the final key of its length $\tau$ with Eve's min-entropy $H_\infty\left(\boldsymbol{X} \mid \boldsymbol{R}, \boldsymbol{C}_{\mathrm{E}}\right)$,

$$H_\infty\left(\boldsymbol{X} \mid \boldsymbol{R}, \boldsymbol{C}_{\mathrm{E}}\right)=-\log_2 \sum_{\left(\boldsymbol{r}, \boldsymbol{c}_{\mathrm{E}}\right) \in \operatorname{Set}\left(\boldsymbol{R}, \boldsymbol{C}_{\mathrm{E}}\right)} \operatorname{Pr}\left(\boldsymbol{r}, \boldsymbol{c}_{\mathrm{E}}\right) \max_{\boldsymbol{x}} \operatorname{Pr}\left(\boldsymbol{x} \mid \boldsymbol{r}, \boldsymbol{c}_{\mathrm{E}}\right). \tag{52}$$

Note that $\boldsymbol{C}_{\mathrm{E}}$ is the ciphertext observed by Eve under the effect of quantum noise and DSR, which corresponds to $\boldsymbol{X}$ for legitimate users.

$$H_\infty\left(\boldsymbol{X} \mid \boldsymbol{R}, \boldsymbol{C}_{\mathrm{E}}\right) \geq \kappa. \tag{53}$$

$$\sum_{\left(\boldsymbol{r}, \boldsymbol{c}_{\mathrm{E}}\right) \in \operatorname{Set}\left(\boldsymbol{R}, \boldsymbol{C}_{\mathrm{E}}\right)} \operatorname{Pr}\left(\boldsymbol{r}, \boldsymbol{c}_{\mathrm{E}}\right)\left|\operatorname{Pr}\left(\boldsymbol{H}\left(\boldsymbol{K}_{\mathrm{R}}\right) \mid \boldsymbol{r}, \boldsymbol{c}_{\mathrm{E}}\right)-2^{-\tau}\right| \leq 2 \varepsilon. \tag{54}$$

$$\varepsilon=\exp\left(\tfrac{1}{2}\left[\tau-H_\infty\left(\boldsymbol{X} \mid \boldsymbol{R}, \boldsymbol{C}_{\mathrm{E}}\right)\right] \ln 2\right). \tag{55}$$

### B. OPTIMUM LEFTOVER HASHING

From (53)–(55), the upper-bound of Eve's average guessing probability on $\boldsymbol{H}(\boldsymbol{K}_{\mathbf{R}})$ is

$$\sum_{\left(\boldsymbol{r}, \boldsymbol{c}_{\mathrm{E}}\right) \in \operatorname{Set}\left(\boldsymbol{R}, \boldsymbol{C}_{\mathrm{E}}\right)} \operatorname{Pr}\left(\boldsymbol{r}, \boldsymbol{c}_{\mathrm{E}}\right) \operatorname{Pr}\left(\boldsymbol{H}\left(\boldsymbol{K}_{\mathrm{R}}\right) \mid \boldsymbol{r}, \boldsymbol{c}_{\mathrm{E}}\right) \leq 2 \varepsilon+2^{-\tau}. \tag{56}$$

The derivation of (56) is shown in Appendix C.

As discussed previously [32], [33], there is an optimum sacrifice amount in LHL as follows.

$$\partial\left(2 \varepsilon+2^{-\tau}\right) / \partial \tau=0 \Rightarrow \tau=\tfrac{1}{3} H_\infty\left(\boldsymbol{X} \mid \boldsymbol{R}, \boldsymbol{C}_{\mathrm{E}}\right). \tag{57}$$

If the final key is shorter than the above amount, it is considered "over-sacrificing," whereas if the final key is longer than the optimal final key, it is considered "under-sacrificing." Hence,

$$\sum_{\left(\boldsymbol{r}, \boldsymbol{c}_{\mathrm{E}}\right) \in \operatorname{Set}\left(\boldsymbol{R}, \boldsymbol{C}_{\mathrm{E}}\right)} \operatorname{Pr}\left(\boldsymbol{r}, \boldsymbol{c}_{\mathrm{E}}\right) \operatorname{Pr}\left(\boldsymbol{H}\left(\boldsymbol{K}_{\mathrm{R}}\right) \mid \boldsymbol{r}, \boldsymbol{c}_{\mathrm{E}}\right) \leq 3 \exp\left[-\tfrac{1}{3} H_\infty\left(\boldsymbol{X} \mid \boldsymbol{R}, \boldsymbol{C}_{\mathrm{E}}\right) \ln 2\right]=3 \times 2^{-\tau} \ll P_{\mathrm{Th}}. \tag{58}$$

The parameter $P_{\mathrm{Th}}$ is the threshold discussed in Section IV.F.

In the case of the Y00 protocol, there is no classical channel to exchange $\boldsymbol{h}$ contrarily to QKDs, in which $\boldsymbol{h}$ is openly known to Eve. However, she would also try to guess the most likely $\boldsymbol{X}$ based on her most confident keys denoted by $\boldsymbol{r}_{\mathrm{E}}$. Then, the following inequality is derived.

$$\begin{aligned} H_\infty\left(\boldsymbol{X} \mid \boldsymbol{R}_{\mathrm{E}}, \boldsymbol{C}_{\mathrm{E}}\right) & \geq-\log_2 \sum_{\boldsymbol{c}_{\mathrm{E}} \in \operatorname{Set}\left(\boldsymbol{C}_{\mathrm{E}}\right)} \operatorname{Pr}\left(\boldsymbol{c}_{\mathrm{E}}\right) \max_{\boldsymbol{r}_{\mathrm{E}}, \boldsymbol{x}} \operatorname{Pr}\left(\boldsymbol{x} \mid \boldsymbol{r}_{\mathrm{E}}, \boldsymbol{c}_{\mathrm{E}}\right) \\ & :=\min_{\boldsymbol{r}_{\mathrm{E}}} H_\infty\left(\boldsymbol{X} \mid \boldsymbol{R}_{\mathrm{E}}, \boldsymbol{C}_{\mathrm{E}}\right) \end{aligned}. \tag{59}$$

Then, instead of (57), the following is the optimal key length.

$$\tau=\tfrac{1}{3} \min_{\boldsymbol{r}_{\mathrm{E}}} H_\infty\left(\boldsymbol{X} \mid \boldsymbol{R}_{\mathrm{E}}, \boldsymbol{C}_{\mathrm{E}}\right). \tag{60}$$

Eve's corresponding guessing probability on $\boldsymbol{H}(\boldsymbol{K}_{\mathbf{R}})$ is

$$\sum_{c_E \in \mathrm{Set}(C_E)} \Pr(c_E)\Pr(H(K_R) \mid r_E, c_E) \le 3\exp\left[-\tfrac{1}{3}\min_{r_E} H_\infty(X \mid R_E, C_E)\ln 2\right] = 3\times 2^{-\tau} \ll P_{\mathrm{Th}}. \quad (61)$$

Therefore, to obtain the valid lengths of the fresh keys for the Y00 protocol, typically $|\boldsymbol{k}_{\mathrm{new}}| + |\Delta\boldsymbol{k}_{\mathrm{new}}| = |\boldsymbol{h}(\boldsymbol{k}_{\mathbf{R}})| = 256$ or 512 bit, $\min_{\boldsymbol{r}} H_\infty(X \mid R, C_E)/3$ must be requested for the final key lengths, while Eve's guessing probability on $\boldsymbol{H}(\boldsymbol{K}_{\mathbf{R}})$ must be suppressed, as suggested previously by (61).

## 6. FUTURE REMARKS

In the key-refreshment process discussed in Section V, Eve may launch so-called "entangling probe attacks" to steal fresh keys as well as the initial keys discussed in Section VI.D of our previous study [27] by preparing her quantum system and then performing joint unitary operations on her system with the signal states between Alice and Bob. Further generalization may be possible that Eve would keep eavesdropping by entangling probe attack during the key-refreshment/initial-key-agreement as well as KPA during the message exchanges after the key-refreshment, which corresponds to coherent attack in QKDs.

The effect of such an attack on the Y00 protocol may be limited; however, evaluation of the strength of such classes of attacks is required. At least, in the key refreshment process, quantum minimax problem in [13], [34], [35] may be helpful; Alice and Bob exchange the keys with a prior probability to minimize Eve's success probability, while Eve derives her optimal measurement operators to maximize her success probability.

In contrast, in the message transmission processes, Eve would not require such a class of attacks because the plaintext is already available to her, while the purpose of the entangling probe attack is to steal the exchanged information in the context of QKDs.

Rather, the critical problem in this study is that the analyses would not give any concrete designs of the Y00 systems because all analyses have done abstractly to find what is the necessary condition to implement ITS Y00 systems. Hence it would neither guarantee whether existing Y00 systems are ITS. However, the study showed the possibility and important parameter for ITS Y00 systems. Some more studies would be required to evaluate the security of the designed systems more easily compared to the security parameter given in this study, which is hard to estimate.

## 7. CONCLUSIONS

This study showed the important security parameter to request the Y00 systems to be ITS and what parameter determines whether the designed Y00 systems are non-ITS or ITS against an attacker who has unlimited computational power with the assistance of quantum memory and unlimitedly long known-plaintext attacks. The analyzed condition is called "collective attacks" or "coherent attacks" in the context of QKD protocols. The conclusions are that Y00 systems remain ITS under certain conditions explicitly provided in this study. Furthermore, this study showed that the attacker's confidence in the shared correct key set increases as time passes. Therefore, a method to refresh the sets of shared keys is proposed using LHL. It had been believed that Y00 protocols are computationally secure. However, this study showed ITS Y00 systems are possible by pointing out what parameters are important to design ITS Y00 systems. To find the above requirement, security analyses were done abstractly. Hence, the critical problem in this study is that the analyses would not give any concrete designs of ITS Y00 systems. Hence, it would not guarantee whether existing Y00 systems are ITS. However, the study showed the possibility and the destination to design ITS Y00 systems. Some more studies would be required to evaluate the security of the designed systems more easily compared to the security parameter given in this study, which is hard to estimate.

## APPENDIX

### *A. DERIVATION OF (10)*

From the equality between conditional and joint entropies,

$$H\left(\boldsymbol{S} \mid \boldsymbol{C}_{\mathrm{E}}, \boldsymbol{X}\right)=H\left(\boldsymbol{S}, \boldsymbol{C}_{\mathrm{E}}, \boldsymbol{X}\right)-H\left(\boldsymbol{C}_{\mathrm{E}}, \boldsymbol{X}\right). \tag{62}$$

An inequality between a conditional entropy and entropy with additional information gives

$$H\left(\boldsymbol{C}_{\mathrm{E}}, \boldsymbol{X}\right) \leq H\left(\boldsymbol{C}_{\mathrm{E}}, \boldsymbol{X}, \boldsymbol{E}\right). \tag{63}$$

The equality is satisfied only when $\boldsymbol{E}$ is a deterministic function of $\boldsymbol{C}_{\mathrm{E}}$ and $\boldsymbol{X}$ irrespective of what $\boldsymbol{S}$ is. Such a situation occur when a sufficient number of bits in $\boldsymbol{S}$ are not hidden under quantum noise enough; then Eve can correct errors by simulation of the system and estimate the error patterns to correct errors using error-correcting code with LFSR, leading to successful FCAs [16], [17].

Eve would obtain $\boldsymbol{S}$ if she could know $\boldsymbol{E}$. Hence,

$$H\left(\boldsymbol{S}, \boldsymbol{C}_{\mathrm{E}}, \boldsymbol{X}, \boldsymbol{E}\right)-H\left(\boldsymbol{C}_{\mathrm{E}}, \boldsymbol{X}, \boldsymbol{E}\right)=H\left(\boldsymbol{S} \mid \boldsymbol{C}_{\mathrm{E}}, \boldsymbol{X}, \boldsymbol{E}\right)=0. \tag{64}$$

Therefore, (10) is derived from (62)–(64).

### *B. DERIVATION OF (29)–(31)*

Consider the following summation with Ω(All) as a set of all patterns of $\{n(\boldsymbol{e} \mid \boldsymbol{r}, \boldsymbol{x})\}$ with the total being $N$. Denote a set of possible signal sequences originating from the shared key stream $\boldsymbol{r}$ as $\Omega(\boldsymbol{r} \mid \boldsymbol{x})$, and its complemental set $\Omega(\boldsymbol{r} \mid \boldsymbol{x})^{\mathrm{C}} := \Omega(\mathrm{All}) - \Omega(\boldsymbol{r} \mid \boldsymbol{x})$.

$$
\begin{aligned}
1 &= \left[\Pr(\boldsymbol{e}_0 \mid \boldsymbol{r},\boldsymbol{x}) + \Pr(\boldsymbol{e} \neq \boldsymbol{e}_0 \mid \boldsymbol{r},\boldsymbol{x})\right]^T \\
&= \sum\nolimits_{\{n(\boldsymbol{e}|\boldsymbol{r},\boldsymbol{x})\}\in\Omega(\mathrm{All})} \binom{N}{n(\boldsymbol{e}_0|\boldsymbol{r},\boldsymbol{x})} \left( \Pr(\boldsymbol{e}_0 \mid \boldsymbol{r},\boldsymbol{x})^{n(\boldsymbol{e}_0|\boldsymbol{r},\boldsymbol{x})} \times \left[\Pr(\boldsymbol{e}_1 \mid \boldsymbol{r},\boldsymbol{x}) + \Pr(\boldsymbol{e} \neq \boldsymbol{e}_0,\boldsymbol{e}_1 \mid \boldsymbol{r},\boldsymbol{x})\right]^{T-n(\boldsymbol{e}_0|\boldsymbol{r},\boldsymbol{x})} \right) \\
&= \sum\nolimits_{\{n(\boldsymbol{e}|\boldsymbol{r},\boldsymbol{x})\}\in\Omega(\mathrm{All})} [N!] \prod\nolimits_{\boldsymbol{e}\in\{0,1\}^{|\boldsymbol{e}|}} \left[n(\boldsymbol{e} \mid \boldsymbol{r},\boldsymbol{x})!\right]^{-1} \Pr(\boldsymbol{e} \mid \boldsymbol{r},\boldsymbol{x})^{n(\boldsymbol{e}|\boldsymbol{r},\boldsymbol{x})}
\end{aligned} . \quad (65)
$$

The last equation is derived from

$$
\binom{N}{n_1}\binom{N-n_1}{n_2}\binom{N-n_1-n_2}{n_3}\cdots\binom{N-\ldots-n_{|\boldsymbol{e}|}}{n_{|\boldsymbol{e}|}} = [N!]\prod\nolimits_{k=1}^{|\boldsymbol{e}|}[n_k!]^{-1} . \quad (66)
$$

Furthermore, note that

$$
\begin{aligned}
1 &= \sum\nolimits_{\{n(\boldsymbol{e}|\boldsymbol{r},\boldsymbol{x})\}\in\Omega(\mathrm{All})} [N!] \prod\nolimits_{\boldsymbol{e}\in\{0,1\}^{|\boldsymbol{e}|}} \left[n(\boldsymbol{e} \mid \boldsymbol{r},\boldsymbol{x})!\right]^{-1} \Pr(\boldsymbol{e} \mid \boldsymbol{r},\boldsymbol{x})^{n(\boldsymbol{e}|\boldsymbol{r},\boldsymbol{x})} \\
&\geq \sum\nolimits_{\{n(\boldsymbol{e}|\boldsymbol{r},\boldsymbol{x})\}\in\Omega(\boldsymbol{r}|\boldsymbol{x})} [N!] \prod\nolimits_{\boldsymbol{e}\in\{0,1\}^{|\boldsymbol{e}|}} \left[n(\boldsymbol{e} \mid \boldsymbol{r},\boldsymbol{x})!\right]^{-1} \Pr(\boldsymbol{e} \mid \boldsymbol{r},\boldsymbol{x})^{n(\boldsymbol{e}|\boldsymbol{r},\boldsymbol{x})} := \Pr(\boldsymbol{r} \mid \boldsymbol{r},\boldsymbol{x})
\end{aligned} . \quad (67)
$$

Hence, $\Pr(\boldsymbol{r} \mid \boldsymbol{x}, \boldsymbol{r})$ is strictly less than unity unless $|\Omega(\boldsymbol{r} \mid \boldsymbol{x})| = |\Omega(\mathrm{All})|$. Then, using $\Omega(\boldsymbol{r} \mid \boldsymbol{x})^{\mathrm{C}}$,

$$
1 - \Pr(\boldsymbol{r} \mid \boldsymbol{r},\boldsymbol{x}) = \sum\nolimits_{\{n(\boldsymbol{e}|\boldsymbol{r},\boldsymbol{x})\}\in\Omega(\boldsymbol{r}|\boldsymbol{x})^{\mathrm{C}}} [N!] \prod\nolimits_{\boldsymbol{e}\in\{0,1\}^{|\boldsymbol{e}|}} \left[n(\boldsymbol{e} \mid \boldsymbol{r},\boldsymbol{x})!\right]^{-1} \Pr(\boldsymbol{e} \mid \boldsymbol{r},\boldsymbol{x})^{n(\boldsymbol{e}|\boldsymbol{r},\boldsymbol{x})} . \quad (68)
$$

Now, using the following inequality,

$$
\begin{aligned}
&\sum\nolimits_{\{n(\boldsymbol{e}|\boldsymbol{r},\boldsymbol{x})\}\in\Omega(\boldsymbol{r}|\boldsymbol{x})^{\mathrm{C}}} [N!] \prod\nolimits_{\boldsymbol{e}\in\{0,1\}^{|\boldsymbol{e}|}} \left[n(\boldsymbol{e} \mid \boldsymbol{r},\boldsymbol{x})!\right]^{-1} \Pr(\boldsymbol{e} \mid \boldsymbol{r},\boldsymbol{x})^{n(\boldsymbol{e}|\boldsymbol{r},\boldsymbol{x})} \\
&\geq \min\nolimits_{\{n(\boldsymbol{e}|\boldsymbol{r},\boldsymbol{x})\}\in\Omega(\boldsymbol{r}|\boldsymbol{x})^{\mathrm{C}}} \prod\nolimits_{\boldsymbol{e}\in\{0,1\}^{|\boldsymbol{e}|}} \Pr(\boldsymbol{e} \mid \boldsymbol{r},\boldsymbol{x})^{n(\boldsymbol{e}|\boldsymbol{r},\boldsymbol{x})} \sum\nolimits_{\{n(\boldsymbol{e}|\boldsymbol{r},\boldsymbol{x})\}\in\Omega(\boldsymbol{r}|\boldsymbol{x})^{\mathrm{C}}} [N!] \prod\nolimits_{\boldsymbol{e}\in\{0,1\}^{|\boldsymbol{e}|}} \left[n(\boldsymbol{e} \mid \boldsymbol{r},\boldsymbol{x})!\right]^{-1}
\end{aligned} . \quad (69)
$$

In contrast, the size of $|\Omega(\boldsymbol{r} \mid \boldsymbol{x})|$ must be $\Pr(\boldsymbol{r})(2M)^T$ because there are $(2M)^T$ patterns of possible detected signal patterns, which suggest $|\Omega(\mathrm{All})| = (2M)^T$. Hence,

$$
\left|\Omega(\boldsymbol{r} \mid \boldsymbol{x})^{\mathrm{C}}\right| := \sum\nolimits_{\{n(\boldsymbol{e}|\boldsymbol{r},\boldsymbol{x})\}\in\Omega(\boldsymbol{r}|\boldsymbol{x})^{\mathrm{C}}} [N!] \prod\nolimits_{\boldsymbol{e}\in\{0,1\}^{|\boldsymbol{e}|}} \left[n(\boldsymbol{e} \mid \boldsymbol{r},\boldsymbol{x})!\right]^{-1} = [1-\Pr(\boldsymbol{r})](2M)^T . \quad (70)
$$

Therefore, (29)–(31) are derived by letting $n(\boldsymbol{e} \mid \boldsymbol{r},\boldsymbol{x}) = N$ for the $\min_{\boldsymbol{e}} \Pr(\boldsymbol{e} \mid \boldsymbol{r},\boldsymbol{x})$ as follows.

$$
\begin{aligned}
&\left(\min\nolimits_{\{n(\boldsymbol{e}|\boldsymbol{r},\boldsymbol{x})\}\in\Omega(\boldsymbol{r}|\boldsymbol{x})^{\mathrm{C}}} \prod\nolimits_{\boldsymbol{e}\in\{0,1\}^{|\boldsymbol{e}|}} \Pr(\boldsymbol{e} \mid \boldsymbol{r},\boldsymbol{x})^{n(\boldsymbol{e}|\boldsymbol{r},\boldsymbol{x})}\right)\left(\sum\nolimits_{\{n(\boldsymbol{e}|\boldsymbol{r},\boldsymbol{x})\}\in\Omega(\boldsymbol{r}|\boldsymbol{x})^{\mathrm{C}}} [N!] \prod\nolimits_{\boldsymbol{e}\in\{0,1\}^{|\boldsymbol{e}|}} \left[n(\boldsymbol{e} \mid \boldsymbol{r},\boldsymbol{x})!\right]^{-1}\right) \\
&= [1-\Pr(\boldsymbol{r})]\left[\min\nolimits_{\boldsymbol{e}\in\{0,1\}^{|\boldsymbol{e}|}} (2M)^{T_{\mathrm{LCM}}} \Pr(\boldsymbol{e} \mid \boldsymbol{r},\boldsymbol{x})\right]^N
\end{aligned} . \quad (71)
$$

### *C. DERIVATION OF (56)*

From LHL (53)–(55), the following decomposition is derived.

$$
\begin{aligned}
2\varepsilon &\geq \sum\nolimits_{(\boldsymbol{r},\boldsymbol{c}_{\mathrm{E}})\in\mathrm{Set}(\boldsymbol{R},\boldsymbol{C}_{\mathrm{E}})} \Pr(\boldsymbol{r},\boldsymbol{x}_{\mathrm{E}}) \left|\Pr(\boldsymbol{H}(\boldsymbol{K}_{\mathrm{R}}) \mid \boldsymbol{r},\boldsymbol{c}_{\mathrm{E}}) - 2^{-\tau}\right| \\
&= \sum\nolimits_{\Pr(\boldsymbol{H}(\boldsymbol{K}_{\mathrm{R}})|\boldsymbol{r},\boldsymbol{c}_{\mathrm{E}})\geq 2^{-\tau}} \Pr(\boldsymbol{r},\boldsymbol{x}_{\mathrm{E}}) \left[\Pr(\boldsymbol{H}(\boldsymbol{K}_{\mathrm{R}}) \mid \boldsymbol{r},\boldsymbol{c}_{\mathrm{E}}) - 2^{-\tau}\right] + \sum\nolimits_{\Pr(\boldsymbol{H}(\boldsymbol{K}_{\mathrm{R}})|\boldsymbol{r},\boldsymbol{c}_{\mathrm{E}})\leq 2^{-\tau}} \Pr(\boldsymbol{r},\boldsymbol{c}_{\mathrm{E}}) \left[2^{-\tau} - \Pr(\boldsymbol{H}(\boldsymbol{K}_{\mathrm{R}}) \mid \boldsymbol{r},\boldsymbol{c}_{\mathrm{E}})\right]
\end{aligned} . \quad (72)
$$

The first term on the last side of (72) shows,

$$2\varepsilon \geq \sum_{\Pr(\boldsymbol{H}(\boldsymbol{K}_{\mathrm{R}})|\boldsymbol{r},\boldsymbol{c}_{\mathrm{E}})\geq 2^{-\tau}} \Pr(\boldsymbol{r},\boldsymbol{c}_{\mathrm{E}})\left[\Pr(\boldsymbol{H}(\boldsymbol{K}_{\mathrm{R}})|\boldsymbol{r},\boldsymbol{c}_{\mathrm{E}})-2^{-\tau}\right]. \tag{73}$$

The second term in the last side of (72) shows

$$\sum_{\Pr(\boldsymbol{H}(\boldsymbol{K}_{\mathrm{R}})|\boldsymbol{r},\boldsymbol{c}_{\mathrm{E}})\leq 2^{-\tau}} \Pr(\boldsymbol{r},\boldsymbol{c}_{\mathrm{E}})\left[2^{-\tau}-\Pr(\boldsymbol{H}(\boldsymbol{K}_{\mathrm{R}})|\boldsymbol{r},\boldsymbol{c}_{\mathrm{E}})\right]\geq 0. \tag{74}$$

Combining (73) and (74), (56) can be derived.

## ACKNOWLEDGMENT

The main concept of this work was described therein when the author was affiliated with Tamagawa University as the previous pre-print was uploaded as arxiv.org/abs/1902.05192. The author especially acknowledges Prof. Osamu Hirota for having offered an opportunity to study the Y00 protocol, quantum information, and quantum optics. The author also acknowledges the staff at Mie University because the author could not have finalized this paper without their support.